\documentclass[3p,12pt, authoryear]{elsarticle}

\usepackage{amssymb}
\usepackage{amsthm}

\usepackage{lineno}
\usepackage{setspace}
\usepackage[colorinlistoftodos,textwidth=100pt]{todonotes}

\usepackage{subfigure}
\usepackage{float}
\usepackage{color}

\newcommand{\ie}{{\it i.e.,~}}
\newcommand{\eg}{{\it e.g.,~}}

\usepackage{xcolor}
\usepackage{natbib}
\setcitestyle{authoryear}
\usepackage{hyperref}
\hypersetup{
    colorlinks=true,
    linkcolor=blue,
    filecolor=magenta,      
    urlcolor=cyan,
}
\newcommand{\mybibstyle}{plainnat}

\usepackage{caption}

\journal{International Journal of Solids and Structures}

\begin{document}

\begin{frontmatter}


\title{Adaptation of the tapered~double~cantilever~beam test for the measurement of fracture energy and its variations with crack speed}

\author[1]{Aditya Vasudevan\fnref{fn1}}
\author[1,2]{Thiago Melo Grabois\fnref{fn1}}
\author[3]{Guilherme Chagas Cordeiro}
\author[4]{St\'{e}phane Morel}
\author[2]{Romildo Dias Toledo Filho}
\author[1]{Laurent Ponson\corref{cor1}}
\ead{laurent.ponson@upmc.fr}

\cortext[cor1]{Corresponding author}
\fntext[fn1]{Equal contribution}

\address[1]{Institut Jean Le Rond d'Alembert, (UMR 7190), Sorbonne Universit\'{e} - CNRS, Paris, France.}
\address[2]{Programa de Engenharia Civil, COPPE, Universidade Federal do Rio de Janeiro, Rio de Janeiro, RJ, Brazil.}
\address[3]{Laborat\'{o}rio de Engenharia Civil, Universidade Estadual do Norte Fluminense Darcy Ribeiro, Campos dos Goytacazes, RJ, Brazil.}
\address[4]{Institut de M\'{e}canique et d'Ing\'{e}nierie (UMR 5295), D\'{e}partement G\'{e}nie Civil et Environnement, Universit\'{e} de Bordeaux, Bordeaux, France.}

\begin{abstract}
In this work we present the design of a new test geometry inspired by the Tapered Double Cantilever Beam (TDCB) specimen that is shown to provide an improved characterization of the fracture properties of brittle solids. First, we show that our new design results in an exponential increase of the specimen compliance with crack length, leading to an extremely stable crack growth during the test. We determine an analytical description of this behavior, 	which provides a simple procedure to extract the fracture energy without depending on finite element calculations. Validation tests are done on polymethylmethacrylate (PMMA) specimens. We use both finite element simulations and our analytical model to interpret the data.  We find a very good agreement between the toughness determined by both methods. The stable nature of crack growth in our improved TDCB specimens results in a precise control of the crack speed.  This feature is employed to go one step further and characterize the variations of toughness with crack speed. We propose an original optimization procedure for the determination of the material parameters characterizing the kinetic law describing the toughness rate dependency. Overall, the approach proposed together with the newly designed test geometry offer unprecedented possibilities for the full and accurate characterization of the fracture behavior of  brittle materials such as rocks, sandstone, mortar etc. 
\end{abstract}


\end{frontmatter}



\section{Introduction}
\label{Intro}

Crack propagation is the main mode of failure of materials under traction. As a result, characterizing the resistance of materials to crack growth is crucial in engineering science for designing safe structures and equipments. Unfortunately, measuring these properties is not straightforward at all. The fracture energy $ G_c$, that provides the amount of energy required to make a crack propagate over a unit surface can be measured by mechanical tests~\citep{Lawn,AtkinsonBook,ASTM}, and more recently, from the statistical analysis of fracture surfaces~\citep{PatentPonson,Auvray}. Contrary to statistical fractography that can achieve a direct measurement of the dissipation rate $G_c$, mechanical test provides an indirect measurement of the fracture energy, through the measurement of the elastic energy release rate $G$, that corresponds to the amount of mechanical energy released by the loaded specimen as the crack propagates over an infinitesimal length, and that is equal to $G_c$ during crack growth. As a result, measuring $G_c$ by mechanical test requires to track both the crack tip position and the elastic energy stored (and so released) by the specimen during the entire test. For that reason, toughness measurement by fracture tests remains complex and prone to experimental errors. In addition, fracture is out-of-equilibrium by nature so that the resistance to failure $G_c$ is rate-dependent, and varies with crack speed. For certain materials, the fracture energy can vary from a factor two or three with the crack propagation velocity ($v$) so this effect cannot be neglected. Recent studies showed that the rate-dependency of fracture energy plays a crucial role in setting the effective toughness of heterogeneous materials~\citep{Chopin}. These rate-dependent effects can be embedded in the so-called {\it{fracture kinetic law}} $G_c(v)$ that provides the variations of the fracture energy with the crack speed $v$. Measuring accurately $G_c(v)$ requires stable fracture both in terms of crack propagation (\ie nearly constant crack speed) and trajectory (straight crack growth). These can be particularly challenging in brittle solids such as glass, rocks or mortar for which crack growth is hard to control for several reasons, then the choice of the geometry of the fracture test that controls the crack stability becomes essential. 

\begin{figure}
    \includegraphics[width=0.45\textwidth]{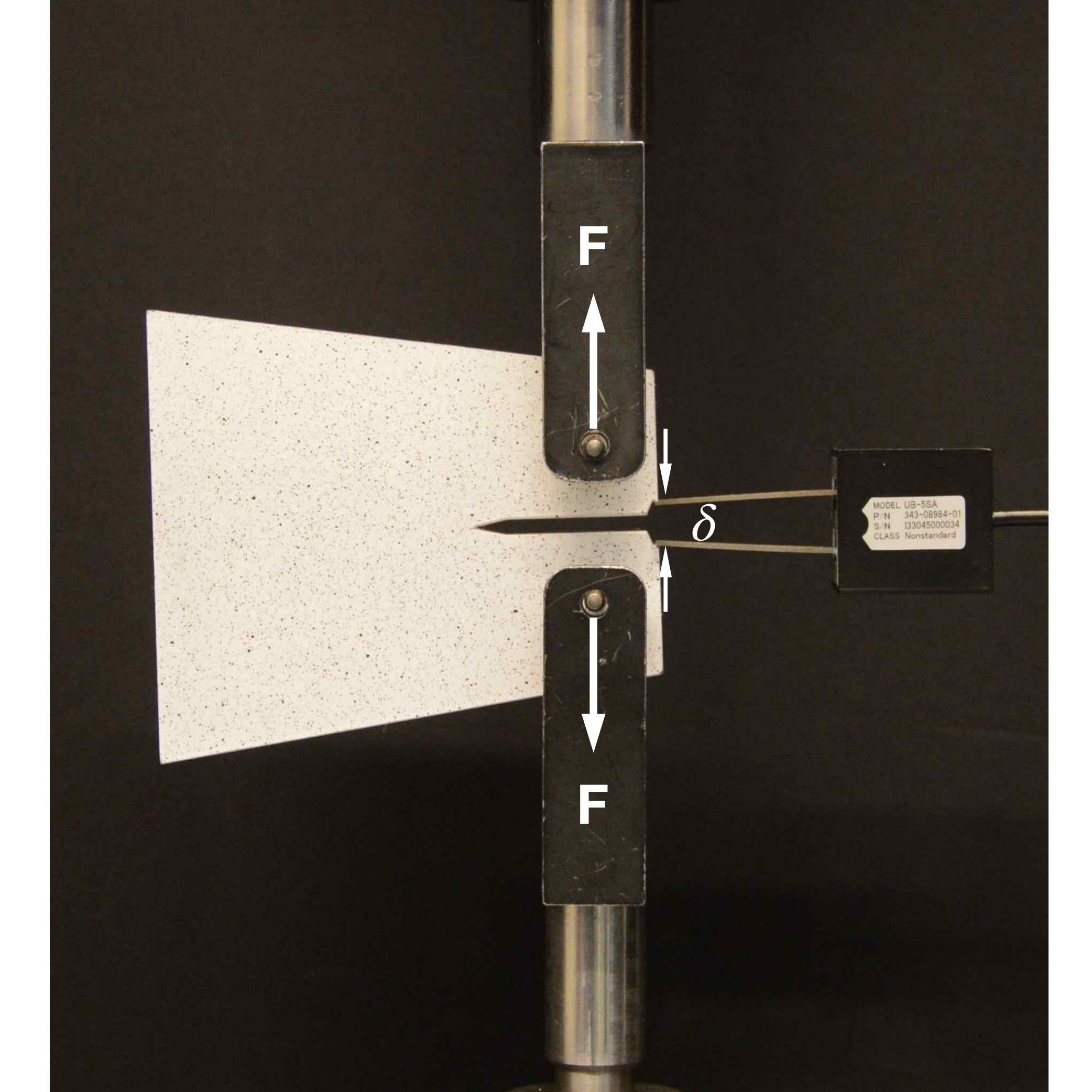}
    \centering
    \caption{The adapted TDCB fracture test geometry mounted on the loading machine.}
    \label{Fig:experimental_setup}
\end{figure}

Different fracture tests have been proposed to handle these challenges. Tests such as the Single Edge Notch Bending (SENB) and Compact Tension (CT) present a short crack propagation length and a rather poor stability of the fracture process~\citep{Sih2}, while the Double Torsion (DT) has a complex crack geometry rendering the interpretation of mechanical data rather complex. Alternatively, the Tapered Double Cantilever Beam (TDCB) test, which has a tapered geometry, enables significant stability of the fracture process and is also easily amenable to DIC measurements~\citep{Grabois}. But so far, it has been mainly used to study adhesively bonded joints~\citep{Marcus,Gallagher,Davalos_1} and quasi-brittle materials like wood~\citep{Morel10} or mortars~\citep{Morel6} that exhibit a pronounced R-curve behavior providing naturally stability to the crack growth process. Here, we take inspiration from these tests, but modify the original TDCB geometry to make it amenable to the study of the failure behavior of brittle solids over a wide range of crack speeds. Through the example of PMMA, we show that the newly proposed fracture test constitutes a rather simple and powerful mean to measure the toughness of brittle solids and its variations with crack speed. 

The classical TDCB geometry uses a straight portion in front of the taper to facilitate the application of the load. Moreover, the ratio of the width to the length of the specimen is generally less than 0.4. We remove these features (see Fig.~\ref{Fig:experimental_setup}) and explore ratios of width to length close to unity which results in an unexpected {\it exponential} variation of the compliance with the crack length and hence an exponential decay of the energy release rate with crack length under fixed imposed displacement. This behavior results in an increased {\it stability} of the crack propagation process. As our data will show, this allows us to measure accurately the fracture energy over a wide range of velocities. We harness this high stability and develop a global optimization procedure allowing to accurately determine the material parameters characterizing the kinetic law, $G_c(v)$, relating toughness and crack speed.

The paper is organized as follows: The first part presents the new design of the TDCB fracture test.  We demonstrate, via numerical simulations, how this new design significantly enhances the stability of crack propagation. We also present a simple analytical description of the mechanical response of the newly designed TDCB specimen from which experimental data can be interpreted simply for measuring toughness. The next section concerns the application of this new fracture test to PMMA specimens allowing us to test the validity of our analytical model and illustrate the practical implementation of our method. The last section is dedicated to the measurement of the fracture kinetic law. Mechanical data obtained from tests achieved at different driving rates are treated using an original optimization method in order to measure the material parameters,  characterizing the kinetic law $G_c(v)$. The application of our method to other materials in other contexts is finally discussed.

\section{Design of the improved TDCB fracture test}

\begin{figure}
    \centering
    \includegraphics[trim = 0 0 0 0, clip = true, width=0.95\textwidth]{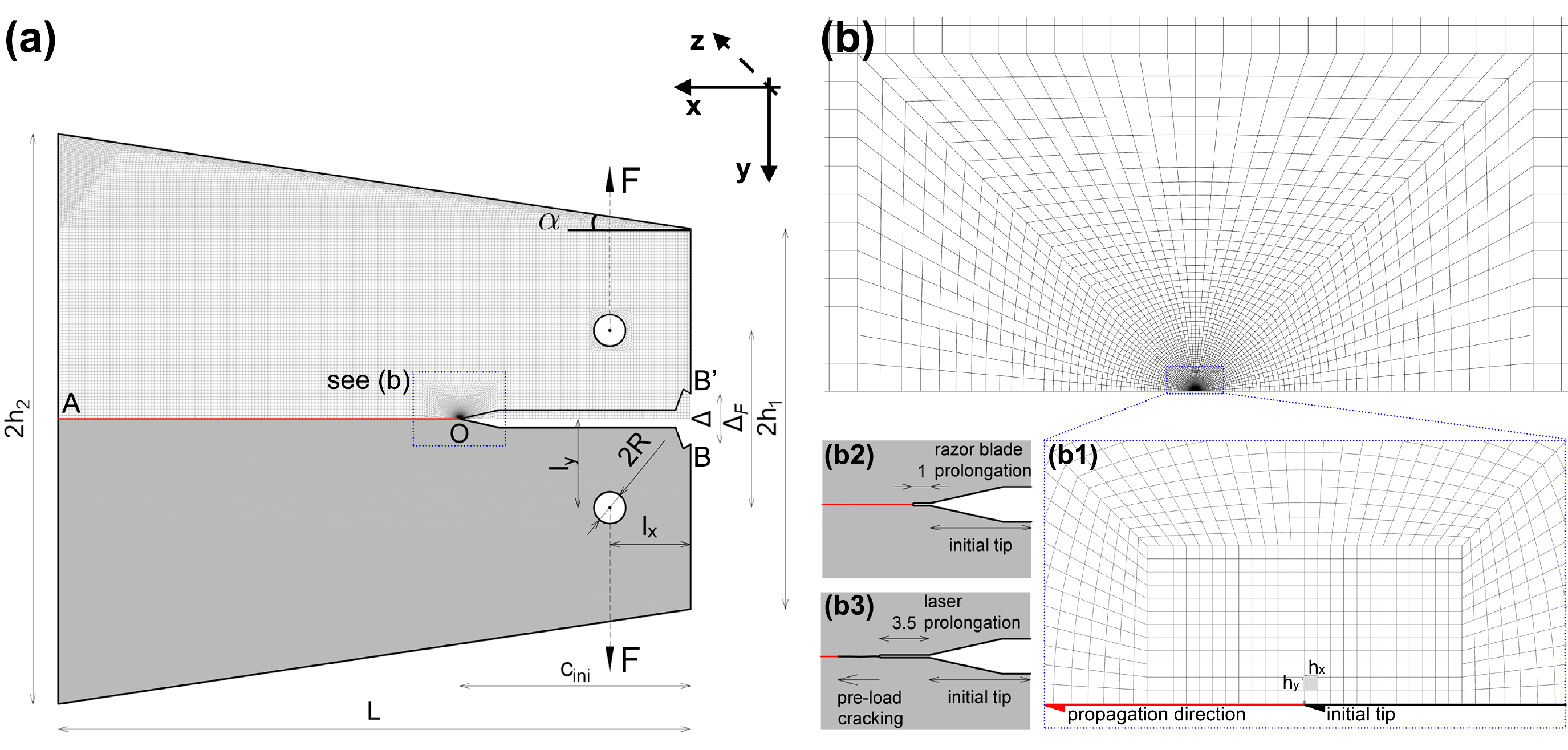}
    \caption{(a) Schematic of the sample geometry with the finite element mesh superposed on the upper half (b) Finite element mesh at the crack tip vicinity. Note the exponential decay of the mesh size as one gets closer to the crack tip as shown in (b1). The mesh used for analysis contains about $ \sim 10^5 $ nodes and $ 2 \times 10^5 $ degrees of freedom. (b2) and (b3) show the experimental methodology employed to prepare the tip of the notch and ensure a smooth crack initiation. In (b2) the crack is sharpened by a razor blade while in (b3) a pre-crack is created to result in a sharp notch.}
    \label{Fig:geometry}
\end{figure}

A schematic of the proposed test geometry is shown in Fig.~\ref{Fig:geometry}(a). The narrow arms of the original design of the TDCB specimen~\citep{Blackman1} have been removed and the height of the sample is now comparable to its length. To measure the fracture properties, we use the compliance method that extracts the crack length by comparing the experimentally measured compliance to a theoretically predicted curve. For slender specimens \--- such as the Double Cantilever Beam (DCB) \--- beam theory can be applied to obtain this theoretical curve, but as our geometry has a large height/length ratio we use finite element (FE) simulations to calculate it. Quadrilateral elements with linear interpolation functions are used and the mesh size in both the $ \mathrm{x} $ and $ \mathrm{y} $ directions are denoted as $ \mathrm{e_x = e_y}$. A typical specimen has dimensions $ \mathrm{h_1} = 120~\mathrm{e_x} $, $ \mathrm{h_2} = 180~\mathrm{e_x} $, $ \mathrm{L} = 400~\mathrm{e_x} $ , $ \mathrm{l_x} = \mathrm{l_y} = 60~\mathrm{e_x} $ and $ \mathrm{R} = 12~\mathrm{e_x} $ that correspond to $ \mathrm{h_1} = 3~\mathrm{cm} $, $ \mathrm{h_2} = 4.5~\mathrm{cm} $, $ \mathrm{L} = 10~\mathrm{cm} $ , $ \mathrm{l_x} = \mathrm{l_y} = 1.5~\mathrm{cm} $ and $ \mathrm{R} = 0.3~\mathrm{cm} $ with $\mathrm{ e_x = e_y} = 250~\mu\mathrm{m}$. We run 2D plane stress calculation assuming that the problem is invariant along the third direction z.

CASTEM, an open-source finite element package developed by CEA, France, is used for our simulations. Exploiting the symmetry in the system, we simulate only the upper half of the sample. A typical mesh configuration used in the analysis is shown in Fig.~\ref{Fig:geometry}. The mesh is coarse in the body with a mesh size $ \rm{e_x} = 250 \rm{\mu m}$, that reduces exponentially to $\rm{e_x} = 1 \times 10^{-11} \rm{m}$ as we reach the tip of the crack as shown in Figure~\ref{Fig:geometry}(b). A unit force is applied at the point P on the hole where the pins apply a tensile force and the displacement is constrained along the Y-direction on the ligament (OA in Fig.\ref{Fig:geometry}(a)) to ensure pure Mode-I fracture. Crack length is increased incrementally and the corresponding elasticity problem is solved under plane stress and linear elastic conditions. 

We extract the displacements $\mathrm{\delta_F}$ and $\mathrm{\delta}$ at the point P of application of the force and at the crack mouth opening (\ie location of the clip gauge in the experimental setup) (B'), respectively (see Fig.~\ref{Fig:geometry}(a)). We define the ratio $ \mathrm{r = \delta_F/\delta} $ that turns out to depend weakly on the crack length and remains close to $ \mathrm{r\simeq 0.8}$. The compliance at these two locations defined as the ratio of the displacement over the force are computed and noted  $ \mathrm{\lambda_F}$ and $\mathrm{\lambda}$ respectively. Note that they are related by the same ratio $r$. All the theoretical calculations are performed using the compliance at the point of application of force. However, as we experimentally measure crack mouth opening displacement (CMOD) at B' using a clip gauge (see Fig.~\ref{Fig:geometry}), the ratio $r$ relates to both quantities. We assume a Poisson's ratio $\nu = 0.37$ that corresponds to PMMA and set the Young's modulus $E$ to unity as it will be subsequently corrected analytically in the expressions of the compliance and elastic energy release rate. Note, however, that the computed value of the compliance $\lambda_\mathrm{F}$ and stress intensity factor $K_\mathrm{I}$ in our 2D simulations depend only slightly on the Poisson's ratio. 

\subsection{Numerical investigation of the mechanical response of the TDCB specimen}
\label{Sec:FEM}

\begin{figure}
    \center{\includegraphics[width = 0.6\textwidth]{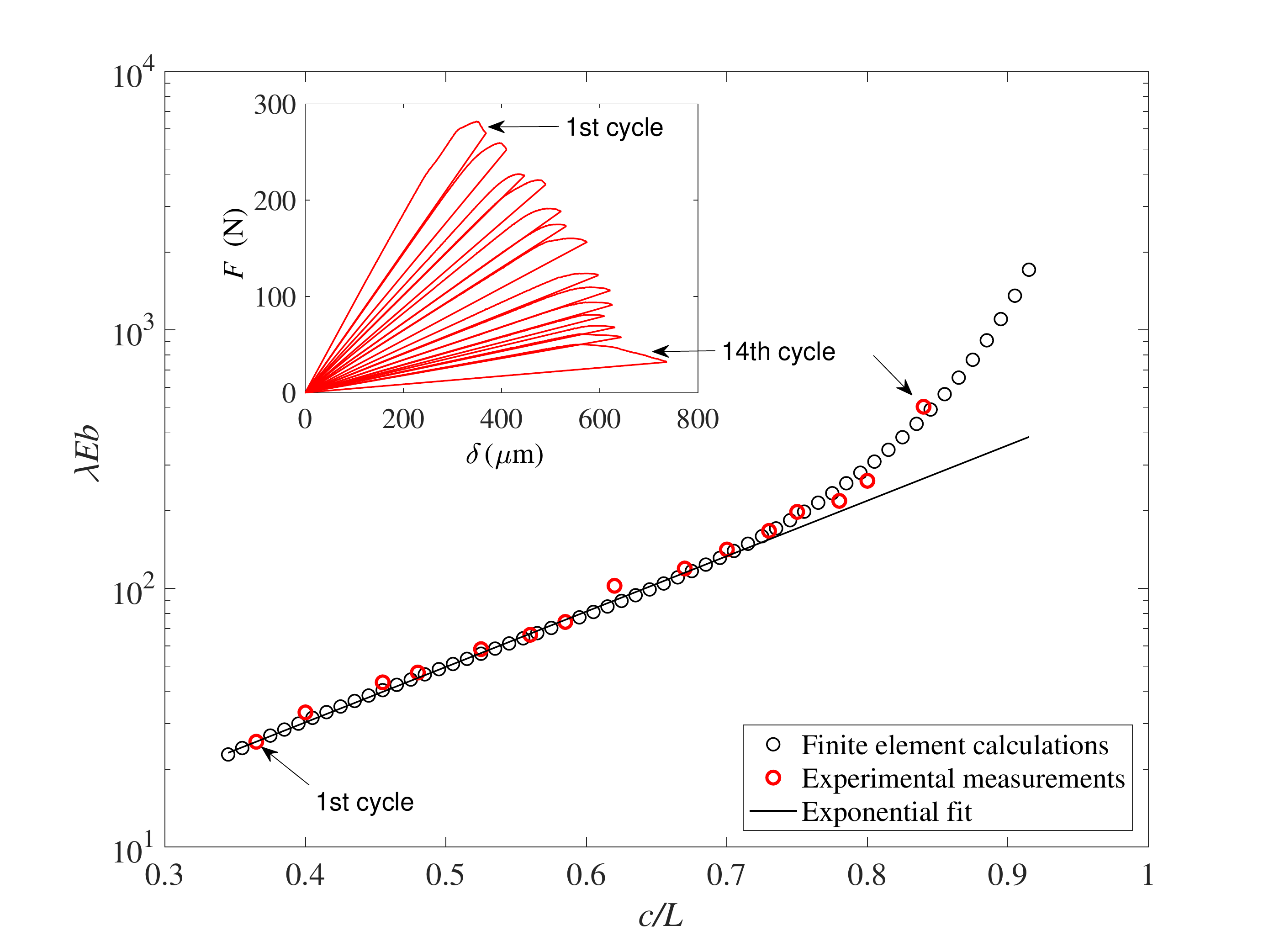}}
    \caption{Normalized compliance as a function of normalized crack length. The black dots are finite element calculations while the red dots are experimental measurements of the compliance that is obtained by 14 cycles of loading-unloading-reloading as shown in the $ F-\delta$ curve in the inset.}
    \label{Fig:compliance_verification}
\end{figure}

A semi-log plot of the non-dimensional compliance obtained from the finite element simulations is shown in Fig.~\ref{Fig:compliance_verification} as a function of the normalized crack length. We see that for a certain range of crack lengths, the dependence of the compliance on the crack length can be described by an exponential function such as (solid line in Fig.~\ref{Fig:compliance_verification}):
\begin{equation}
    \lambda_F = \frac{\lambda_0}{Eb} e^{c/c_0}
    \label{Eq:lambda_fit}
\end{equation}
where $E$ is the Young's modulus, $b$ the thickness of the specimen while $\mathrm{\lambda_0}$ and $\mathrm{c_0}$ are parameters of the exponential fit that depend on the geometry of the specimen. The compliance follows an exponential increase in a mid-range of crack lengths from $\mathrm{ L/3 < c < 2L/3}$, where $L$ is the specimen length along the propagation direction. Once the crack reaches the vicinity of the free surface, the ligament becomes small and the compliance increases faster than an exponential and it deviates away from the exponential fit. This exponential variation of the compliance provides immense stability to crack propagation.

We then perform a systematic analysis on the scaling of the parameters of the fit ($c_0$ and $ \lambda_0$) by performing simulations for different geometries. We choose two different geometries that have the same taper angle $ \alpha \sim 8.5^\mathrm{o}$, but different lengths $ l_x$ (\ie that corresponds to the distance from the hole, where a pin apply the force, to the boundary) and $ h_1$ (\ie that represents the height of the specimen).  We also vary the length of the specimen $ L $. The first geometry has $ l_x = 90 e_x$ and $ h_1 = 160 e_y$, while the second geometry has $l_x = 90 e_x$ and $ h_1 = 120 e_y$. The inset of Fig.~\ref{Fig_FE} shows the parameter $ c_0 $ extracted from the fit for all the simulations. It turns out that  $ c_0$ depends only on $L$ and varies as $ \mathrm{c_0 = 0.389 L - 0.0229 }$, where $\mathrm{c_0}$ and $L$ are in meter. We find the other fitting parameter, the amplitude $\lambda_0$ of the exponential law, scales as $ \mathrm{c_0^2/(h_1 l_x) }$. This scaling is shown from the collapse of the curves in Fig.~\ref{Fig_FE} corresponding to different values of $ c_0 $, $ h_1 $ and $ l_x $. This behavior remains valid in the range $\mathrm{60 e_x \leqslant l_x \leqslant 90~e_x}$, $ \mathrm{120~e_x \leqslant h_1 \leqslant 160~e_x}$ and $400~e_x\leqslant L \leqslant 1200~e_x$. Using this scaling to normalize the axis in Fig.~\ref{Fig_FE}, different geometries lead to the same exponential regime in some limited range of crack lengths. In this range, the compliance can then be approximated by

\begin{equation}
    \lambda_F \simeq \frac{1}{Eb} \beta \frac{c_0^2}{h_1l_x} e^{c/c_0}
    \label{Eq:compliance_normalized}
\end{equation}
where $\beta~\simeq~3.0$ is a constant obtained from the numerical fit. This expression is represented by the solid line in Fig.~\ref{Fig_FE}. This results in a simple formula for the variation of compliance for different geometries, provided  the $8.5^\mathrm{o}$ tapered angle that remains the same. We can now predict the mechanical energy release rate $G$ that will be subsequently required to interpret the experimental data. $G$ can be deduced from the compliance formula also referred to as the Irwin-Keis equation~\citep{Lawn} by the relation: 

\begin{figure}
    \includegraphics[width=0.8\columnwidth]{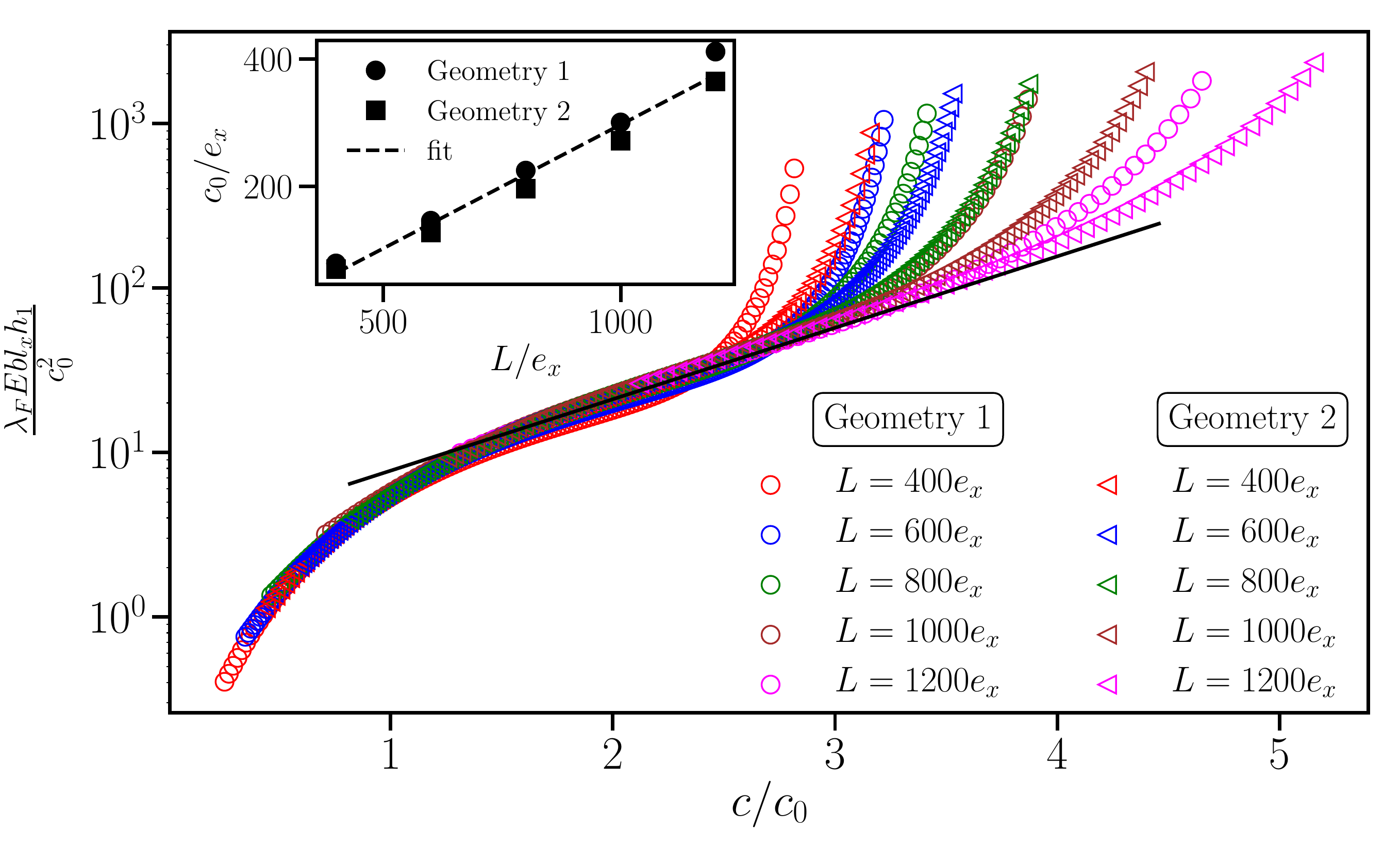}
    \centering
    \caption{Variation of the non-dimensional specimen compliance with the non-dimensional crack length for different specimen length and geometry. The simulations are carried out for two different geometries with the same tapering angle $\mathrm{ \alpha }$ but with different $\mathrm{ l_x }$ and $ \mathrm{h_1} $. The geometry number 1 is represented by circles and has dimensions $ \mathrm{l_x = 90 e_x}$ and $\mathrm{ h_1 = 160 e_y} $ while geometry number 2 is represented by triangles with dimensions $\mathrm{ l_x = 60 e_x} $ and $ \mathrm{h_1 = 120 e_y}$. The solid line shows the exponential fit of the collapsed curve and corresponds to Eq.~\ref{Eq:compliance_normalized}.}
    \label{Fig_FE}
\end{figure}

\begin{equation}
    G = \frac{F^2}{2b} \frac{d\lambda_F}{dc} = F^2 g
    \label{Eq:comp_form}
\end{equation}
where $g$ that corresponds to the geometry dependent part of the elastic energy release rate is given by 
\begin{equation}
    g =\frac{1}{2b} \frac{d\lambda_F}{dc}
    \label{eq:g1}
\end{equation}
Substituting Eq.~\ref{Eq:compliance_normalized} in Eq.~\ref{eq:g1} one obtains its expression as a function of the geometry of the specimen: 
\begin{equation}
    g = \frac{1}{2Eb^2}\frac{\beta c_0}{h_1l_x}e^{c/c_0}
    \label{eq:g2}
\end{equation}

\begin{figure}
    \includegraphics[width=3 in]{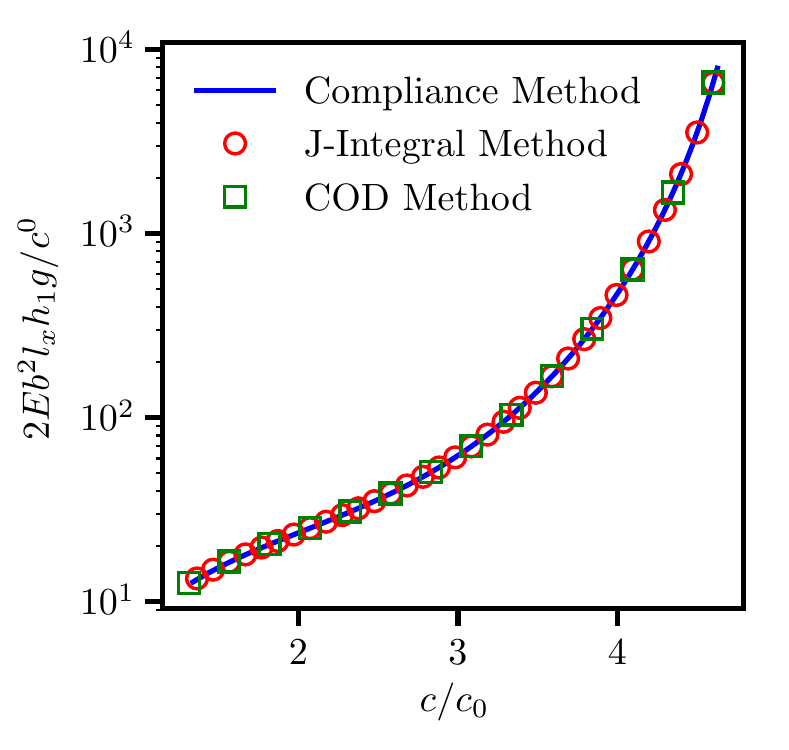}
    \centering
    \caption{Variation of non-dimensional elastic energy release rate $g$ as a function of normalized crack length computed from finite element simulations using three different methods. The simulations correspond to a specimen with $\mathrm{ l_x = 90~e_x }$, $ \mathrm{L = 400~e_x}$ and $\mathrm{ h_1 = 160~e_x}$. }
    \label{Fig_g}
\end{figure}

Our motivation to define $g$ is that it depends only on the specimen geometry, so $G$ can be decomposed into the product of the square of the force applied with a geometry dependent function. As a result, for a unit force, $G = g$ as measured in our simulations where a unit force is imposed. Taking inspiration from Eq.~\ref{eq:g2}, we plot $g$ on a semi-log scale as a function of $c/c_0$ after normalization by $c_0/(2Eb^2h_1l_x)$. The value of $g$ is here obtained by three independent methods: (i) The first method consists in using Eq.~\ref{eq:g1} with the derivative of the compliance computed numerically using the finite element solution (see \eg Fig.~\ref{Fig_FE}); (ii) the second method uses the J-Integral that computes energy flow to the tip of the crack through a closed contour integral around the crack tip~\citep{Rice3}; (iii) and the final and third method relies on the crack opening displacement. From linear elastic fracture mechanics, we know that the crack opening profile follows $u(x) \sim K_I \sqrt{x}$, so the stress intensity factor, $K_I$, is obtained by fitting the computed opening profile $u(x)$. This is done in a logarithmic representation using a linear fit of slope 1/2. The energy release rate is then obtained from  Irwin's relation $G = K_I^2/E$. These methods are employed for different crack lengths and they give very similar results 
as shown in Fig.~\ref{Fig_g}.

As the compliance is known, we can now go from force constant (dead-weight) loading used in the simulation to displacement constant (fixed-grips) loading used in the experiments. To convert the previous expression into fixed grip loading, we must replace $F$ in Eq.~\ref{Eq:comp_form} by $\delta_F / \lambda_F$. Using $r = \delta_F/\delta$ and the expression of the compliance $\lambda_F$ given by Eq.~\ref{Eq:lambda_fit}, one obtains
\begin{equation}
    G = \frac{\delta_F^2}{\lambda_F^2} g= \delta^2 \frac{E r^2e^\frac{-c}{c_0}}{2 \lambda_0 c_0}.
    \label{eq:G_fixed_delta}
\end{equation}
We observe that G decreases exponentially fast with crack length under fixed imposed displacement $\delta $. This is a key property of the modified TDCB test that allows us to explore different velocities and fracture energy. For sake of comparison, let us now consider classical fracture tests like DCB, DT and CT. In DCB and classic TDCB, the compliance increases as the cube of crack length so from Eq.~\ref{eq:G_fixed_delta}, the energy release rate decays as one over fourth power of crack length~\citep{Sih}. In DT test, the compliance varies linearly with crack length and the energy release rate varies inversely to the square of crack length. To have a stable and controlled crack growth, it is decisive that the energy release rate decreases as fast as possible with crack length. The exponential decrease achieved by the TDCB specimens is faster than the one obtained by these classical tests that display only a polynomial decrease, thus ensuring an extremely stable crack growth as illustrated in the following.

\section{Experimental Results}
\label{sec:ER}

We use a transparent thermoplastic, the so-called Polymethylmethacrylate (PMMA) or Plexiglas, as a model brittle solid. Samples are obtained by machining $8~\mathrm{mm}$ thick PMMA sheets through laser cutting. Fracture tests are then carried out under uniaxial tension using a Shimadzu (model AG-Xplus) universal testing machine of $10~kN$ maximum load capacity. In this setup, a force gauge of $1~kN$ load capacity measures the tensile forces applied to the specimen by the two steel pins located in the sample holes and a clip gauge measures the displacements between the lips of the crack (see Fig.~\ref{Fig:experimental_setup}). Experiments are controlled by the clip gauge, imposing a constant crack opening speed. In order to ensure a stable crack propagation and avoid overloading the sample at initiation, we employed two different methodologies to sharpen the V-shape notch machined during the cutting process that provided satisfying results (see Fig.~\ref{Fig:geometry}(b2) and (b3)): (i) we use a razor blade of 0.2 mm thickness to produce propagation over approximately $1~\mathrm{mm}$ and; (ii) after laser cutting a slit of 3.5 mm we pre-load the sample at a crack mouth opening rate of 0.5~$\mathrm{\mu m/s}$ until crack initiates. The sample is then unloaded and the new initial crack length is recorded.

\begin{figure}[H] 
    \centering
    \includegraphics[width=0.45\textwidth, trim = 2mm 2mm 2mm 2mm, clip= true]{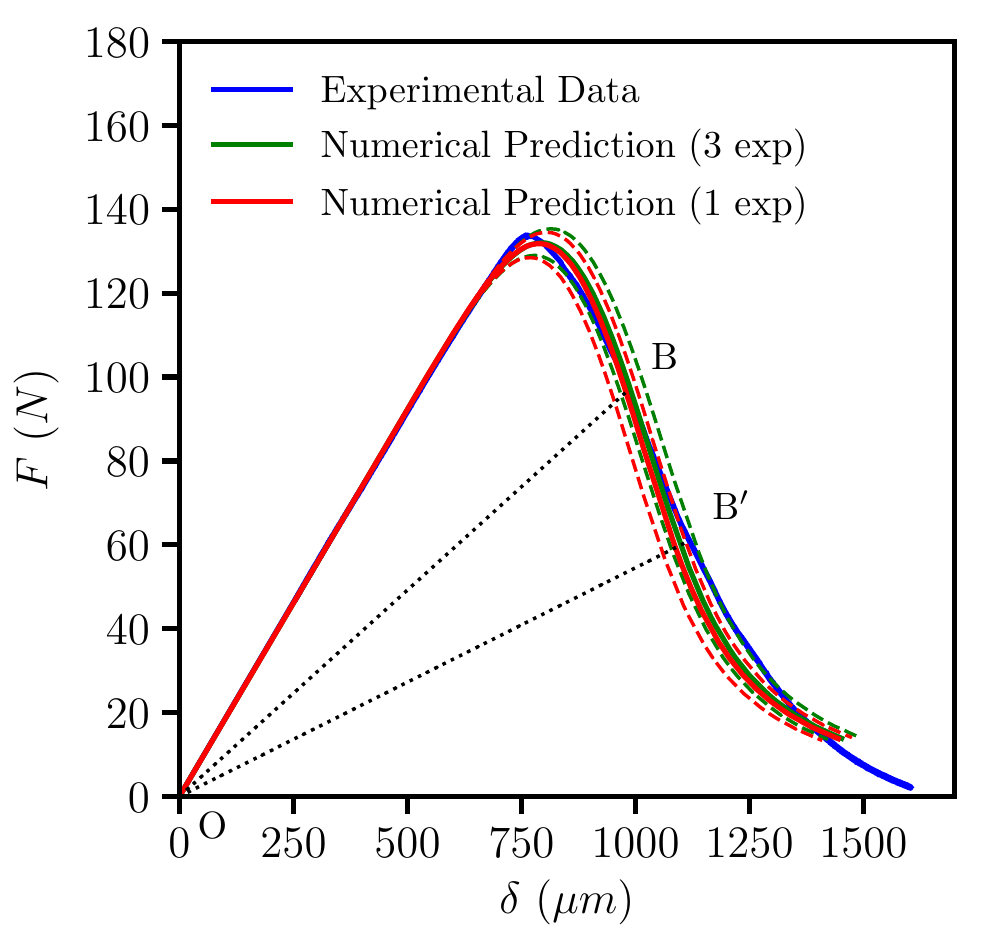}
    \caption{Typical load-crack opening displacement ($F-\delta$) curve for a single test under mode I tensile fracture in a TDCB specimen of PMMA. The blue curve is the experimentally measured curve, while the green and the red curve are theoretically predicted by assuming a kinetic law $G_c(v)$ (see Section~\ref{Sec:kinetic-law}) that is obtained by fitting for three or one experiment respectively. The dotted lines of respective color are the error-bars in the theoretical prediction.}
    \label{Fig_F_delta}
\end{figure}

Fig.~\ref{Fig_F_delta} shows a typical Force~{\it vs.}~displacement curve for a test that is driven with a clip gauge opening rate $\dot{\delta} = d\delta/dt = 2.5~\mu m.s^{-1}$.

\subsection{Experimental validation of the compliance method}
\label{Sec:compliance_verification}

Before measuring the fracture properties of the material, we first test our approach by measuring experimentally the compliance as a function of crack length and comparing it to to our simulation results (see~Fig.~\ref{Fig:compliance_verification}). The experimental compliance is obtained by unloading and reloading the sample at some fixed crack length that is measured independently through optical means. As the sample is fully unloaded, we do not observe any residual displacement, thus validating the linear elastic approach (see inset of Fig. \ref{Fig:compliance_verification}). The slope of the force-displacement curve provides the compliance value after each measurement, we propagate the crack over a small distance to measure the new compliance value for a longer crack. Figure~\ref{Fig:compliance_verification} shows that the experimental and theoretical compliance agree very well. 

\subsection{Measurement of crack length and crack speed}
\label{Sec:c_t_v_t}

\begin{figure}[H]
    \centering
    \begin{tabular}{ccc}
    \subfigure[\label{fig:c-t}]{\includegraphics[trim = 2mm 1mm 2mm 1mm,clip,width=0.31\linewidth]{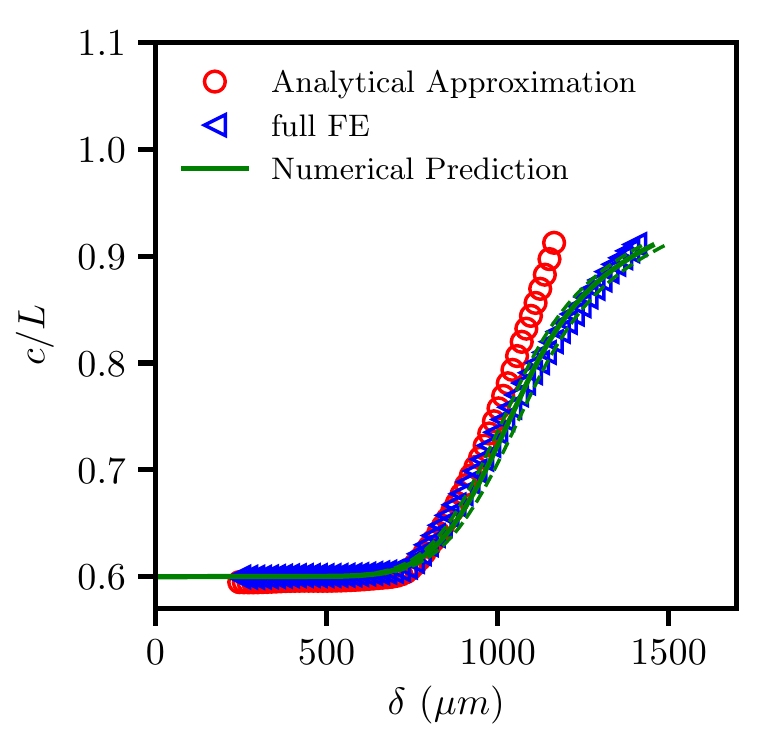}}
    \subfigure[\label{fig:v-c}]{\includegraphics[trim = 2mm 1mm 5mm 1mm,clip,width=0.31\linewidth]{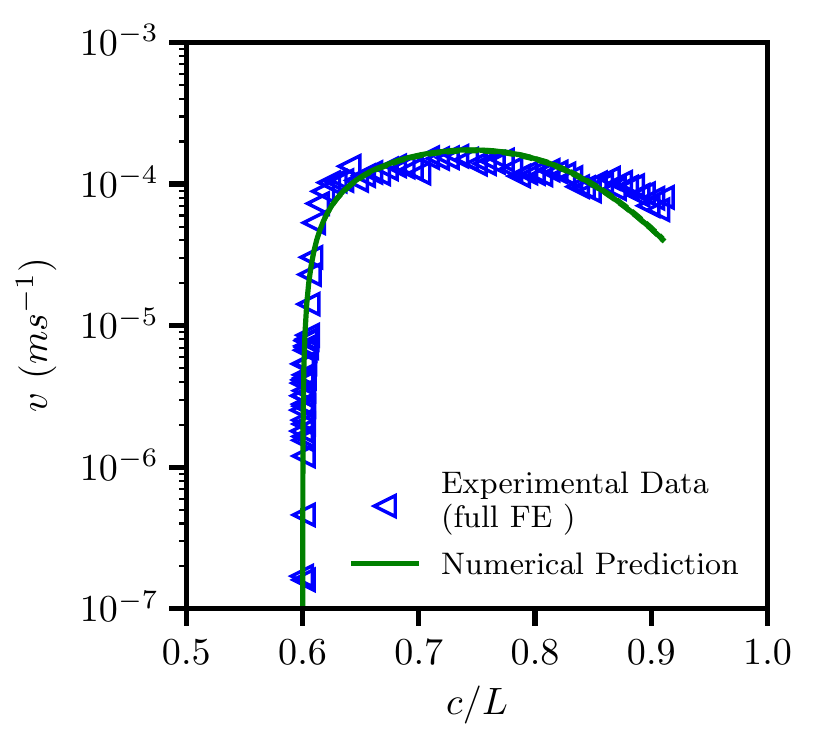}}
    \subfigure[\label{fig:Gc-c}]{\includegraphics[trim = 2mm 1mm 2mm 1mm,clip,width=0.31\linewidth]{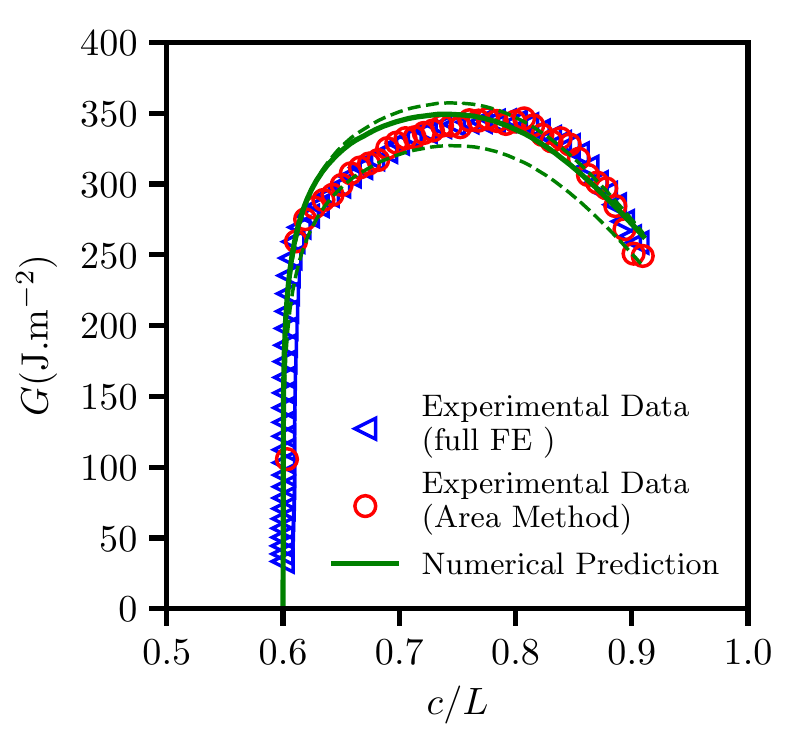}}
    \end{tabular}
    \caption{Experimental measurement of the crack length $c$, its velocity $v$ and energy release rate $G$ computed. (a) The crack length is extracted from the experimental compliance value and comparison either with values computed from FE or with the analytical formula of Eq.~\ref{Eq:compliance_normalized}; (b) measured velocity variations as a function of normalized crack length using the compliance method based on finite element calculations; while (c) Elastic energy release rate {\it vs.} the normalized crack length obtained from the compliance based on finite element calculations (in blue) and the area method (in red) described in Sec.~\ref{Sec:G_c}. Finally, the numerical predictions derived from the kinetic law $G_c(v)$ with the parameters $G_{c0} = 2000 \pm 100$ and $\gamma = 0.2 \pm 0.01$ determined from the optimization method presented in Sec.~\ref{Sec:kinetic-law} is represented in green, with the dotted green line providing the uncertainty propagated from the error bar on the parameters values $G_{c0}$ and $\gamma$.}
    \label{fig:G-v-c}
\end{figure}

The crack tip position $c(t)$ is inferred from the comparison of the experimentally measured compliance $\lambda(t) = \delta/F$ with the one $\lambda^\mathrm{FE}(c)$ computed from finite element simulations. The crack length normalized by the full length L of the sample is plotted in Fig.~\ref{fig:c-t} as a function of the imposed crack mouth opening displacement $\delta$. Along with the use of the full FE results, we also plot the crack length predicted from the analytical formula of Eq.~(\ref{eq:G_fixed_delta}) and observe a good agreement until $c/L \simeq 0.7$, as expected from the limited range of validity of the analytical formula, $L/3 \lesssim c \lesssim 2L/3$. The crack growth velocity $\displaystyle{v(t)=\frac{dc}{dt}}$ is then computed and its variations is shown in Fig.~\ref{fig:v-c}. We observe a fairly constant crack speed in agreement with the finite element study that predicts very stable crack growth.

\subsection{Measurement of the elastic energy release rate}
\label{Sec:G_c}

Once crack length is known, we then have two methods that can be used to measure the elastic energy release rate, $G$. The first method relies on the estimation of the energy dissipated for an incremental crack advance using the force-displacement response. Assume that in Fig.~\ref{Fig_F_delta} in B the crack length is equal to $c$ and that after the time $\Delta t$, it is equal to $c + \Delta c$. The mechanical energy release rate is then given by the energy dissipated during this time divided by the area of the newly created fracture surface. The energy dissipated during this time interval $\Delta t$ is equal to the area OBB'O (see Fig.~\ref{Fig_F_delta}) and the newly created fracture surface area is $\Delta c \, b$ where $b$ is the width of the specimen~\citep{morel7}. Thus, $G(t)$ is given by
\begin{equation}
    G(t) = \frac{E^d(t)}{ \Delta c(t) b}
    \label{eq:G-t}
\end{equation}
where $E^d(t)$ is the energy dissipated that corresponds to the area OBB'O in Fig.~\ref{Fig_F_delta} after replacing $\delta$ by $\delta_F$ using the ratio $r$. In the second method, we use directly the finite element predictions of Eq.~\ref{Eq:comp_form} derived in section~\ref{Sec:FEM}. The energy release rate obtained by both methods is shown in Fig.~\ref{fig:Gc-c} presenting very similar results. So from now on, Eq.~\ref{Eq:comp_form} is used to measure the energy release rate $G$ in our experiments. 

\subsection{Prediction of the average crack speed}
\label{sec:CS}

Our theoretical description of crack growth in TDCB specimens is used to predict the characteristic crack speed $v_\mathrm{m}$ as a function of the imposed crack opening rate~$\dot{\delta}$. Assuming that the fracture energy values remains uniform during a test, the equation of motion $G = G_\mathrm{c}$ together with the expression~\ref{eq:G_fixed_delta} of the elastic energy release rate provides
\begin{equation}
    v_m = r\sqrt{\frac{2 E c_0}{\lambda_0 G_c}} e^{\frac{-c}{2c_0}} \dot{\delta}
    \label{eq:v_m_deltadot}
\end{equation}
after derivation with respect to time. First, we notice that the crack speed decreases exponentially with crack length. However, as the characteristic decay length $2c_0 \simeq 0.8 L$ is relatively large, the range of crack speeds explored during a single test is relatively modest, a feature that can be noticed on Fig.~\ref{fig:G-v-c}(b). Furthermore, $c$ can be taken equal to a the characteristic crack length, typically $c \simeq 60$ mm, thus providing the characteristic crack speed  explored during the test proportional to the imposed crack opening rate $\dot\delta$. 

\begin{figure}[H]
    \centering
    \includegraphics[trim = 0 0 0 13mm, clip,width=0.5\linewidth]{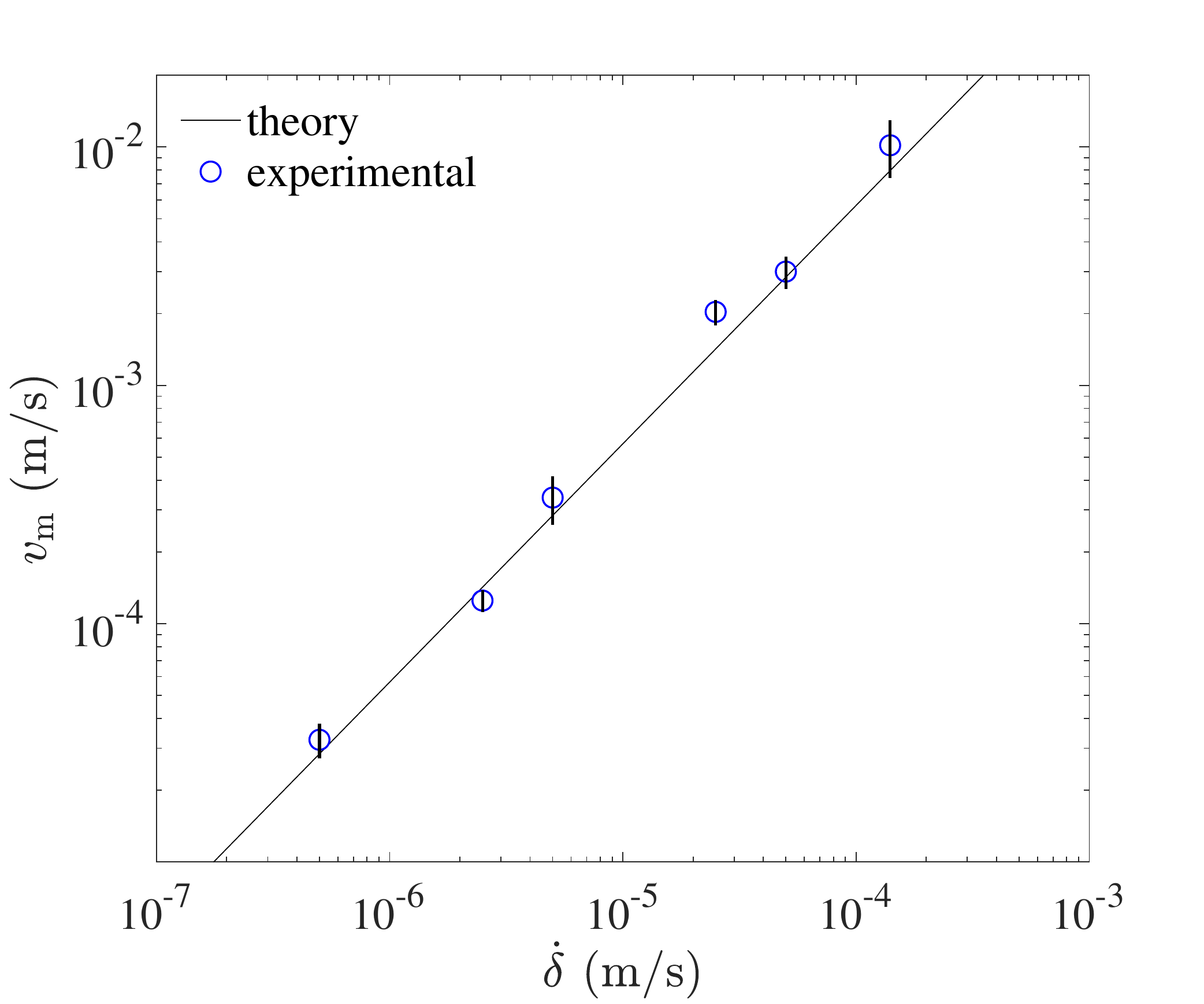}
    \caption{Log-log representation of the average crack speed during the TDCB fracture test as a function of the imposed crack opening rate. The straight line corresponds to the prediction of Eq.~\ref{eq:v_m_deltadot} where the parameters $c_0 = 21~\mathrm{mm}$, $\lambda_0 = 4.81$, and $r = 0.8$ are measured from finite element simulations. The value $E = 1.82~\mathrm{GPa}$ is the PMMA Young's modulus. The adopted characteristic value of the PMMA fracture energy is $G_\mathrm{c} = 380~\mathrm{J.m^{-2}}$.}
    \label{fig:vm-deltadot}
\end{figure}

Figure~\ref{fig:vm-deltadot} shows the variations of the average experimental crack speed for different loading rates in the range $0.5~\mu\mathrm{m}.s^{-1} \leq \dot{\delta} \leq 100~\mu\mathrm{m}.s^{-1}$ imposed during the experiments. Its comparison with the analytical prediction of Eq.~\ref{eq:v_m_deltadot} \--- displayed in a logarithmic representation \--- shows an excellent agreement, reflecting that the crack speed explored experimentally can be finely tuned by changing the loading rate imposed to the specimen. In particular, the crack velocity is about fifty times larger than the loading rate for PMMA material parameters in this new TDCB fracture test configuration. Thus, the application of the analytical formula~\ref{eq:v_m_deltadot} enables to select some desired crack velocity by choosing the value of the crack opening rate.

\section{Measurement of the kinetic law $G_\mathrm{c}(v)$}
\label{Sec:Gc-v}

\begin{figure}[H]
    \includegraphics[width=0.7\textwidth]{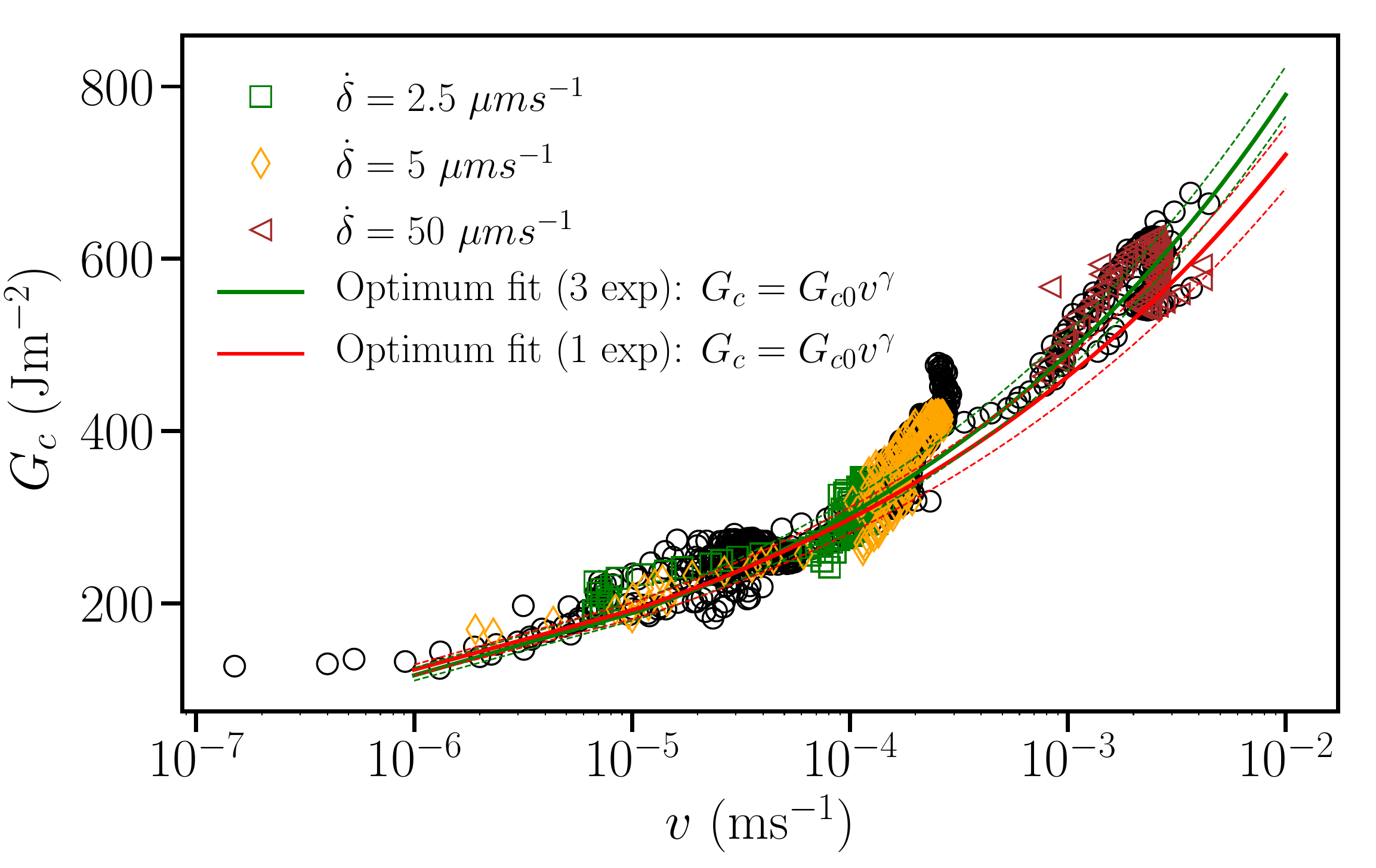}
    \centering
    \caption{Diagram of the variations of the fracture energy with crack speed. All the experimental data are in black circles while the three colored symbols highlight the experimental results used for the optimization procedure, fitting to the kinetic law $G_c(v) = G_{c0} v ^\gamma$. The red and green lines show the fit whether it has considered one or three experiments, respectively. The dashed lines represent the error bars of the fit.}
    \label{fig:G-v} 
\end{figure}

In polymeric materials like PMMA, slow crack propagation proceeds in two steps: the polymeric chains are first gradually aligned until full elongation and then failure takes place~\citep{Berry}. This elongation phase is a rate-dependent process and so depends on the velocity at which the crack propagates. As a result, the fracture energy is also a function of the cracks velocity~\citep{Green}. The so-called $ G_c-v $ curve, also called the fracture kinetic law, is a material property that can be described by a power law in polymeric materials~\citep{Maugis3}. The kinetic law obtained from compiling all our experimental data is provided in Fig.~\ref{fig:G-v}. Interestingly just few experiments are sufficient to capture the kinetic law over a rather extended range of crack speeds. This can be achieved since one single test (for example see the test shown in green in Fig.~\ref{fig:G-v} that corresponds to $\dot{\delta} = 2.5~\mu\mathrm{m.s^{-1}}$) explores a range of crack speed varying over about a factor three.

Here, around six TDCB fracture tests at each imposed crack opening rate $\dot\delta$ were required to characterize the rate-dependency of the fracture energy. In the following section, we describe a methodology that can achieve a similar accuracy, but from much less fracture tests.  
 
%
%

\section{Quantitative characterization of the kinetic law $G_c(v)$ from a limited number of fracture tests}
\label{Sec:kinetic-law}

\subsection{Prediction of the mechanical response of the modified TDCB specimens from the kinetic law}

We first show here how the rate dependency of the fracture energy can be included into our description of the failure behavior of TDCB specimens. For polymers, the fracture energy is generally found to depend rather strongly on crack speed~\citep{Maugis3} which is consistent with our experimental observations. To describe this variation, a two parameter kinetic law of the form
\begin{equation}
    G_c (v) = G_{c0} \, v^\gamma 
    \label{Eq:kinetic_law}
\end{equation}
can be suitably used. The determination of the kinetic law of a material has many interests as it allows to make predictions on the lifetime of a structure or it can be used to infer the microscopic failure mechanisms taking place at the crack tip vicinity during crack growth (like stress corrosion, etc.). However, the experimental determination of the material parameters (like here $G_{c0}$ and $\gamma$) that characterize it is generally very challenging. We here propose a methodology that allows for an improved determination of these material parameters from a limited number of tests. For illustrating the method, we consider the kinetic law of Eq. \ref{Eq:kinetic_law}, and seek to determine that material parameters $G_{c0}$ and $\gamma$.

Let's assume first some values of these two material parameters and predict the corresponding force-displacement response of the modified TDCB specimen under some prescribed opening rate $ \dot{\delta}$. We start with the Griffith's criterion that reads as
\begin{equation}
    G = G_c(v) \Longrightarrow \frac{1}{2b} \left ( \frac{\delta (t)}{\lambda^{FE}[c(t)]} \right ) \left . \frac{d \lambda^{FE}}{d c} \right |_{c(t)} = G_\mathrm{c} [\dot{c}(t)]
    \label{Eq:c_t_PDE}
\end{equation}
where $\delta (t) = \dot{\delta} \, t$ and $G_\mathrm{c} [\dot{c}(t)]$ is given by Eq.~\ref{Eq:kinetic_law}. Equation~\ref{Eq:c_t_PDE} is a nonlinear first order differential equation where the crack length variation $ c(t) $ is unknown. This equation can be solved numerically using an explicit scheme with the initial condition $c(t=0) = c_\mathrm{ini}$, where $c_\mathrm{ini}$ is the crack length at initiation. The force-displacement response is then reconstructed using the parametric functions $ \delta (t) = \dot{\delta} \, t$ and $ F = \dot{\delta} \,t/\lambda^{FE}[c(t)]$, which can be compared with the experimental force-displacement curve as shown in Fig.~\ref{Fig_F_delta}. Comparing the predicted and the experimental mechanical response of the TDCB specimen, one can thus determine the material parameters involved in the kinetic law of Eq.~\ref{Eq:kinetic_law} as described in the following section.

\subsection{Global optimization procedure for the accurate determination of the kinetic law}
\label{sec:globopt}

\begin{figure}[H]
\includegraphics[width=\textwidth]{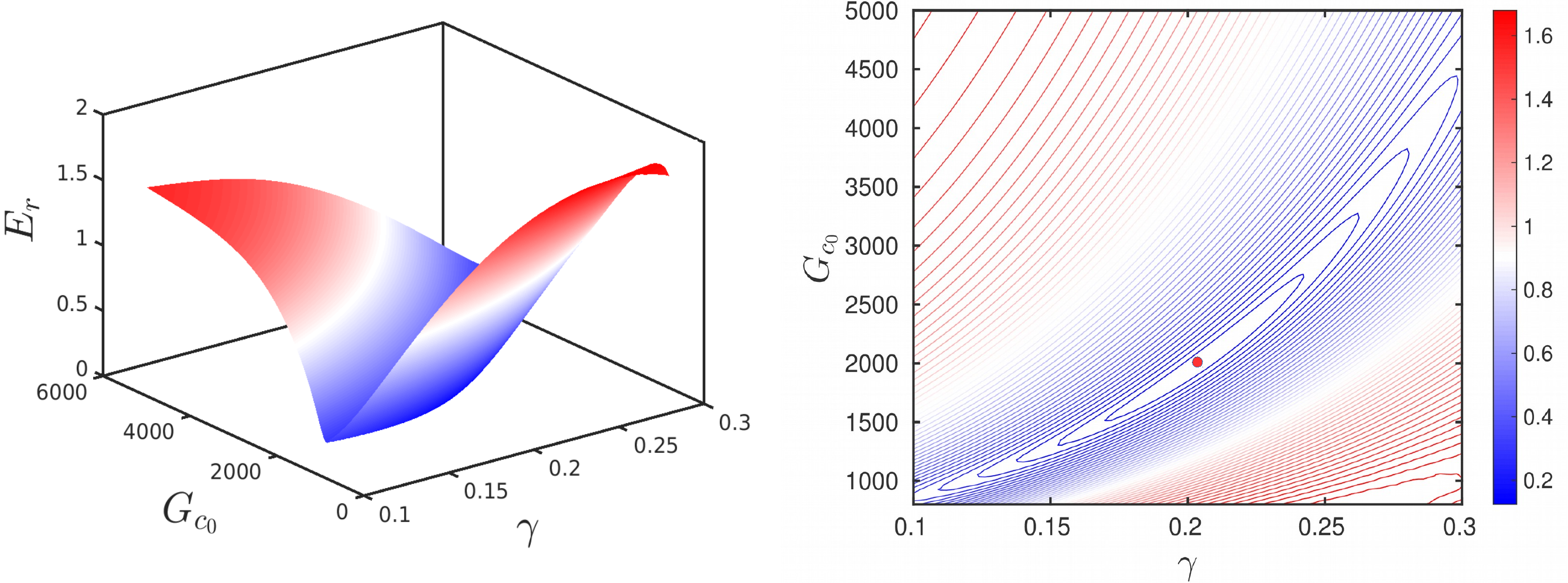}
\centering
\caption{Error on the prediction of the force-displacement response of the modified TDCB specimen as a function of the parameters $ G_\mathrm{c}^0 $ and $ \gamma $ involved in the kinetic law of Eq. \ref{Eq:kinetic_law}.}
\label{Fig:error_plot} 
\end{figure}

To carry out the determination of the kinetic law, we choose three experiments performed at three different loading rates that are represented in color in Fig.~\ref{fig:G-v}. The three experiments are chosen so that a large range of velocities are covered, resulting in an improved determination of the kinetic law. For each experiment, we assume some values of the parameters $G_\mathrm{c}^0$ and $\gamma$ and compute the error
\begin{equation}
    \epsilon_r = \sqrt{\sum_{\delta_i} \left[ \frac{F_\mathrm{test}(\delta_i) - F_\mathrm{exp}(\delta_i)}{F_\mathrm{exp}}\right]^2}
\end{equation}
between the numerically predicted force-displacement curve $F_\mathrm{test}(\delta)$ (following the procedure described in the previous section) and the experimentally measured response $ F_\mathrm{exp}(\delta)$. The procedure is then applied for multiple experiments and the cumulative error for all the experiments is subsequently defined as
\begin{equation}
    E_r = \sqrt[]{\sum_{k=1}^{n} \frac{\epsilon_k^2}{n}}
\label{eq:cumerror}
\end{equation}
where $n$ is the total number of experiments. Figure~\ref{Fig:error_plot} shows a 2D map of the error $E_r$ in the parametric plane $(\gamma,G_{c0})$ for $ n = 3$. The minimum of the error, that corresponds to the best match between the theoretically predicted mechanical response and the experimentally measured one, is found for $G_{c0} = 2000$ and $ \gamma = 0.20$. The uncertainty of both parameters is determined by one percent deviation of the forces predicted from the optimum fit. We find $ \delta G_{c0} \sim \pm 100 $ and $ \delta \gamma \sim \pm 0.01$ and the error bars are plotted in dotted lines in Fig.~\ref{fig:G-v}. In addition, we push the optimization procedure by considering only one experiment (the one that corresponds to $\dot{\delta} = 2.5~\mu \mathrm{m}.{s}^{-1}$ in Fig.~\ref{fig:G-v}) and we measure $G_{c0} = 1750$ and $\gamma = 0.192$ and is plotted in red in Fig.~\ref{fig:G-v} while the forces are plotted in Fig.~\ref{Fig_F_delta}. We find that with this geometry a single experiment could be sufficient to measure the kinetic law, though using at least three experiments that drives cracks at different velocities yields better results as it spans over a broader range. 

Once the kinetic law has been determined, one can predict all the other features of the fracture test like $c(t)$, $v(c)$ and $G(c)$. The prediction of each of these curves is shown in solid line in Fig.~\ref{fig:G-v-c}. They are in very good agreement with the experimentally measured behavior.
\section{Conclusion}
\label{sec:conclusion}

This work demonstrated how an improved TDCB test geometry enhanced the determination of the fracture properties of brittle materials. We proposed a new fracture test inspired from the tapered double cantilever beam specimen that ensures steady crack growth and allows for the precise measurement of the fracture energy. In particular, the numerical results show that the elastic energy release rate decays exponentially fast with crack length, providing extreme stability to the fracture process. Moreover, the experimentally measured compliance and crack length data were in agreement to the simulations, which validate the designed approach.      

The improved TDCB geometry was used in experiments at a macroscopic scale to determine fracture properties (\eg crack tip positions, crack velocities, fracture energy) of PMMA samples. We showed the ability of the experimental configuration to calculate the sought parameters by correlating the experimental force and displacement data to FE results. In addition, data from different driven load velocity tests enabled to build a relationship between the average crack velocity and the imposed opening rate (\ie $v_\mathrm{m} \sim \dot{\delta}$), allowing to explore any desired crack speed. 

The proposed fracture test combined with a numerical resolution of the linear elastic fracture mechanics problem was harnessed to accurately determine the full kinetic law $G_{c}(v)$, characterizing the variations of toughness with a large range of crack speeds. As a result, this law permits us to predict the sought fracture test features such as $c(t)$, $v(c)$, and $G(c)$. Overall, we endorse that this improved TDCB methodology overcomes the limitation of other fracture test geometries by allowing a simple and accurate determination of the material toughness and its variations with crack speed. Another route, which is worth further investigating to validate the approach, is to consider different brittle solids such as rocks, glasses, and cement-based materials.   

\section*{Acknowledgement}
LP and AV acknowledge the support of the City of Paris through the Emergence Program. This study was financed in part by the Coordena\c{c}\~{a}o de Aperfei\c{c}oamento de Pessoal de N\'{i}vel Superior - Brasil (CAPES) -
Finance Code 001. TG's stay at UPMC was supported by the Doctoral Sandwich Program Abroad (PDSE fellowship) of CAPES, grant $\#99999.009821/2014-07$. TG, GC, and RTF also acknowledge the financial support of the Brazilian agencies FAPERJ and CNPq to this investigation.

\bibliographystyle{\mybibstyle}
\bibliography{bibfracture}

\end{document}